\documentclass[traditabstract]{aa_short} % for the abstract without structuration 
\usepackage{graphicx}
\usepackage{txfonts}
\usepackage{natbib}
\usepackage{tabularx}
\bibpunct{(}{)}{;}{a}{}{,}

\title{The VIMOS Public Extragalactic Survey (VIPERS)}
\subtitle{\bf First Release of spectra}

\author{
P. Franzetti \inst{1}
\and B. Garilli \inst{1}
\and L.~Guzzo\inst{2,3}
\and A.~Marchetti\inst{1} 
\and M.~Scodeggio\inst{1} 
on behalf of the VIPERS Team
}
\authorrunning{P. Franzetti et al.}
\titlerunning{VIPERS spectra release}   
\institute{
INAF - Istituto di Astrofisica Spaziale e Fisica Cosmica Milano, via Bassini 15,
20133 Milano, Italy%1
\and INAF - Osservatorio Astronomico di Brera, Via Brera 28, 20122 Milano, via
E. Bianchi 46, 23807 Merate, Italy %2
\and Dipartimento di Fisica, Universit\`a di Milano-Bicocca, P.zza della Scienza
3, I-20126 Milano, Italy %3
}

\begin{document}

\abstract
  % context heading (optional)
 {We release the spectra for the more than 57000 objects presented in the First VIPERS Data Release. 
For each object we distribute the 
observed, wavelength and flux calibrated spectrum, as well as {\it cleaned} spectra, 
where artifacts due to fringing are removed. We also provide the sky and noise spectrum 
and the 2D spectrum. Data can be downloaded from {\tt http://vipers.inaf.it}.}

\keywords{Galaxies: distances and redshifts -- Galaxies: statistics --
  Cosmology: observations --
  Astronomical databases: Catalogues}
\maketitle

\section{Public Release of Spectra}
{The VIMOS Public Extragalactic Redshift Survey (VIPERS;
\citealt{vipers_main}) has been designed to collect $\sim 10^5$ redshifts
to the same depth of VVDS-Wide \citep{vvds_wide} and zCOSMOS \citep{zcosmos_10k} surveys
 ($i_{\rm AB}<22.5$), but over a
significantly larger volume and with high sampling (${\rm
  \sim 40\%}$ ).   
The general aim of the project is to build a sample of the global galaxy population 
that matches in several respects those available locally ($z<0.2$) from the  2dFGRS \citep{2df} and SDSS  \citep{sdss}
projects, thus allowing combined evolutionary studies of both clustering and galaxy physical properties, 
on a comparable statistical footing. Building upon the experience and results of previous 
VIMOS surveys  VIPERS 
arguably provides the most detailed and representative picture
to date of the whole galaxy population and its large-scale structures, when the
Universe was about half its current age. A full description of the survey is given in \citet{vipers_main}.\\
In September 2013, we have released the first
two thirds of the survey (PDR-1, \citet {vipers_PDR}), 
comprising redshifts and redshift quality flag for 57204 sources. 
We are now releasing the 1D and 2D spectra of such objects.\\
Data are distributed through the public VIPERS site, 
{\tt http://vipers.inaf.it}, and can be retrieved either as a single tar file 
containing the full release, or
via the VIPERS Database, described in \citet {vipers_PDR}. 
Publication through the Virtual Observatory will be done in the near future. \\
For each observed object, we provide the spectrum in form of a FITS binary table, 
similar to the format used for SDSS distribution. Files are named using the object IAU identifier. 
Each table contains the following columns:
\begin{itemize}
\item {\bf Wavelength}, column name {\it WAVES}, expressed in $\AA$
\item {\bf Observed spectrum}, column name {\it FLUXES}: the monodimensional spectrum as resulting from the reduction pipeline, wavelength calibrated and corrected for the instrument sensitivity function. The absolute flux calibration has been obtained by normalizing the spectrum to the i band photometric magnitude provided by CFHTLS. Units are {\it $erg/cm^2/s/\AA$}. 
\item {\bf Cleaned spectrum}, column name {\it EDIT}: to further clean the spectra from fringing correction residuals, as well as poor subtraction of strong sky lines and/or zero orders, we have applied a combination of manual cleaning and automatic cleaning and reconstruction, this last performed 
using a PCA based approach (details on this procedure will be given in a forthcoming dedicated paper). The cleaned spectrum is the result of such procedure. Units are {\it $erg/cm^2/s/\AA$}. 
\item {\bf Noise spectrum}, column name {\it NOISE}, the noise spectrum computed during the reduction as described in \citet {vipers_PDR}.Units are {\it $erg/cm^2/s/\AA$}. 
\item {\bf Sky spectrum}, column name {\it SKY}: the sky spectrum of each slit, in counts.
\item {\bf Spectrum mask}, column name {\it MASK}: monodimensional binary mask indicating 
the pixels where the cleaning and reconstruction has been applied: 1 for cleaned and reconstructed pixels.
\end{itemize}
In the header of each file, we report the VIPERS ID of the object, its Right Ascension and declination, the selection $i_{\rm AB}$ magnitude, the spectroscopic redshift and reliability flag, the normalization applied and the exposure time. In a separate file, we also provide the wavelength calibrated 2D image of the slit containing the object. These files are named as the monodimensional spectra with suffix 2D, and the header contains the same information as the 1D spectra file (redshift, flag, coordinates, etc..). \\
When making use of this data set, acknowledge should be given by referring to \citet{vipers_main} and \citet {vipers_PDR}.
}
\bibliographystyle{aa}
\bibliography{VIPERS_spectraPublicRelease}
\end{document}